# Electron heating in a current-driven turbulence as a result of nonlinear interaction of electron- and ion-acoustic waves

Jian Chen[1, 2, 3], Alexander V. Khrabrov[3], Igor D. Kaganovich[3], and He-Ping Li[2]

[1]Sino-French Institute of Nuclear Engineering and Technology, Sun Yat-sen University, Zhuhai 519082, P. R. China

[2]Department of Engineering Physics, Tsinghua University, Beijing 100084, P. R. China

[3]Princeton Plasma Physics Laboratory, Princeton NJ 08543, USA

We study electron heating in collisionless current-driven turbulence due to the nonlinear interactions between electron- and ion-acoustic waves. PIC simulation results show that due to a large difference between the electron and ion mean velocities the Buneman instability excites large-amplitude ion-acoustic waves, which strongly modifies the electron velocity distribution function, leading to a secondary instability that generates fast electron-acoustic waves; and in this process, a giant electron hole is ultimately created. This giant electron hole is responsible for strong electron heating due to phase mixing. The numerical simulation results are consistent with the previous observations and provide insight into the key processes responsible for electron heating and the generation of nonlinear waves in a collisionless current-driven instability.

Current-carrying collisionless plasmas exhibit excitation of many nonlinear waves and complex turbulence. Current-driven instability develops in such a system once the relative drift velocity between the electrons and ions, $v_D$, exceeds a critical value (instability is strongest when $v_D$ exceeds the electron thermal velocity $v_{Te}$) [1]. The instability reaches a nonlinear level and transforms into turbulence that is responsible for several important transport phenomena including anomalous resistivity and electron heating [2-4], both of which have been observed in plasmas used in a range of applications, including the tokamak startup [5, 6], hollow cathode discharges [2-4, 7, 8], space plasmas [9-13], etc. Large efforts have been devoted to better understanding the correlation between the abrupt change of the macroscopic parameters (i.e., the electron temperature) and microscopic turbulence physics.

In warm current-carrying plasmas, the ion-acoustic waves (IAWs) are generated due to the onset of the Buneman instability [14]. Nonlinear processes such as the formation of double layers and phase-space vortices were also identified [15-20]. Besides that, fast and high-frequency electron waves are also observed in both simulations and experiments [9-11, 21-23]. Specifically, cluster spacecraft measurements observed slow electron holes propagating in Earth's magnetotail, which are likely generated by the Buneman instability and responsible for the electron and ion heating [9]. Kinetic simulations by Che *et al*. [11] further have proven that in the current sheet the Buneman instability grows and traps slow electrons while the remaining fast electrons can drive two-stream instability and excite electron waves, as a result fast electrons transfer momentum and energy to the ions and slow electrons. Similar phenomenon of closely- coupled development of the Buneman instability and electron two-stream instability was also observed in the plasma devices utilized for the electric propulsion, and are responsible for the anomalous electron transport in the plasma plume area of devices [7, 23]. Recently, Magnetospheric Multiscale (MMS) mission measured the electron velocity distribution function (EVDF) in the magnetopause reconnection region, which had a plateau region; this observation was explained by generation of large-amplitude ion and electron waves that can create the plateaus at EVDF around their respective resonant velocity [10]. In analysis of Ref.[10], the amplitude of ion waves was used to calculate the plateau width; and the results show that the plateaus created by the ion and electron waves are separate (namely, the ion and electron waves weak interact with each other). However, in our study we observe that large-amplitude solitary ion waves can be generated and these waves may be strongly coupled with electron waves because the plateaus created by the ion and electron waves overlap. *Understanding these coupling effects between the ion and electron waves and their influence on turbulent electron heating is the motivation of this study.*

We performed 1D spatial, 3D velocity particle-in-cell simulations which can capture the nonlinear wave coupling process [24, 25]. We observed a complex interation between IAWs and electron-acoustic waves (EAWs) which are both self-consistently formed in the simulations. Large-amplitude IAWs are created by the Buneman instability and modify EVDF by trapping low-energy electrons in the potential well of the wave. The modified EVDF is subject to a bump-on-tail instability that produces large amplitude EAWs and ultimately evolves into a single solitary wave, sometimes called electron hole or electron solitary wave (ESW), in which electrons are rapidly heated due to the phase mixing.

Our simulations start from a homogeneous plasma with warm ions and electrons with a large relative drift velocity. The mass ratio $m_i/m_e$=1836, plasma density $n$=10$^{17}$ m$^{-3}$, electron and ion temperature $T_e$=$T_i$=7.0 eV, and $v_D$=$v_{Te}$=1.56×10$^6$ m/s are used as the initial condition. Periodic boundaries are used: the particles moving out from one side are re-injected back from the other side. The length of the simulation domain is $L$=1000$\lambda_{De}$ and the spatial step is $\Delta x$=1/3$\lambda_{De}$ (where $\lambda_{De}$=6.3×10$^{-5}$ m is the electron Debye length). A constant electric field $E_0$=5.0 V/m is imposed. This setup describes an inductive tokamak startup. Because we only focus on the electrostatic waves parallel to the magnetic field, the magnetic field is neglected in our 1D simulations [5]. The time step is $\Delta t$=6.0×10$^{-12}$ s ($\omega_e \Delta t$ =0.1). Test runs without an external electric field are performed and successfully reproduce the stability boundary given by linear theory [shown in the Supplementary Material (Session A) for this letter].

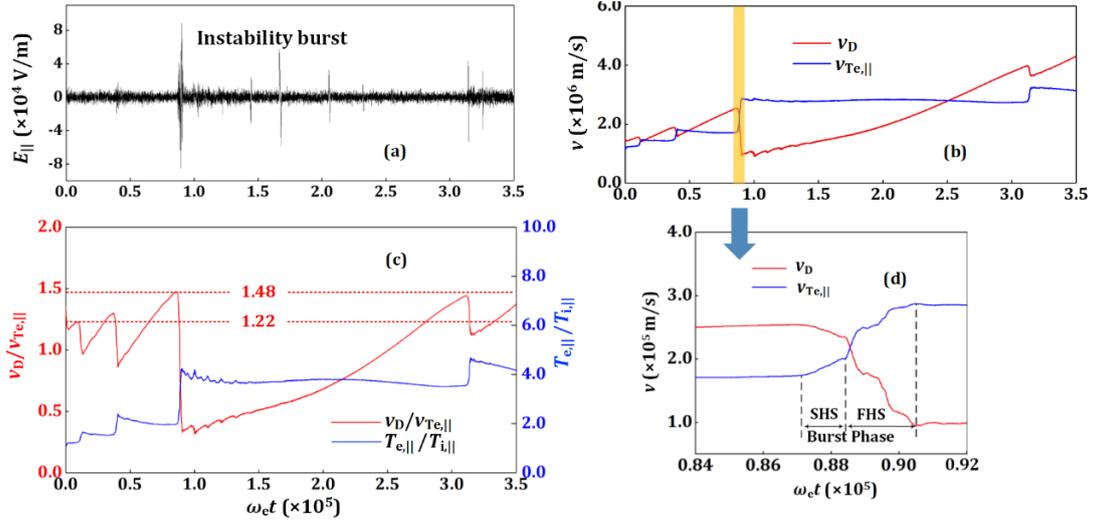

Fig. 1: Evolution of (a) the electric field $E_\parallel$ at $x=158\lambda_{De}$; (b) the electron drift velocity $v_D$ and the thermal velocity $v_{Te,\parallel}$; (c) the velocity ratio $v_D/v_{Te,\parallel}$ and the temperature ratio $T_{e,\parallel}/T_{i,\parallel}$; (d) zoom-in view of the evolution of the velocity $v_D$ and $v_{Te}$ during the burst phase [the orange zoom-in window in (b)], where SHS and FHS represent the slow and fast electron heating stages, respectively. Here, $T_{e,\parallel}$ and $T_{i,\parallel}$ are defined as $T_{e,\parallel} = \sum \frac{1}{2} m_e \left(v_{e,\parallel} - \bar{v}_{e,\parallel}\right)^2$ and $T_{i,\parallel} = \sum \frac{1}{2} m_i \left(v_{i,\parallel} - \bar{v}_{i,\parallel}\right)^2$, where $v_{e,\parallel}$ and $v_{i,\parallel}$ are the parallel (to electric field) velocity component of the particle velocity, $\bar{v}_{e,\parallel}$ and $\bar{v}_{i,\parallel}$ are the average particle velocity. The thermal velocity $v_{Te,\parallel}$ is defined as $v_{Te,\parallel}=\sqrt{2kT_{e,\parallel}/m_e}$ (thus, at initial moment for a Maxwellian EVDF, $v_{Te}=\sqrt{2}v_{Te,\parallel}$). The simulation parameters are: $m_i/m_e=1836$, $n=10^{17}$ m$^{-3}$, $T_e=T_i=7.0$ eV, and $v_D=v_{Te}=1.56\times10^6$ m/s.

Fig. 1(a) show time evolution of the electric field $E_\parallel(t)$ at $x=158\lambda_{De}$ (the subscript "$\parallel$" represents the direction aligned with the external electron field). Several pulses of spiky fluctuations are evident in Fig.1, corresponding to intermittent instability bursts. A similar time evolution of $E_\parallel$ has already been observed in many satellite measurements [9, 10, 26], indicating that the current-driven instability may have an intermittent nature in space as well. Accordingly, the time evolution of $v_D$ and $v_{Te,\parallel}$ show a quasi-periodic pattern. Before and after the burst phase, $v_D$ increases nearly linearly with time whereas $v_{Te,\parallel}$ is nearly constant. However, when the instability burst happens, $v_{Te,\parallel}$ rises rapidly and $v_D$ dramatically decreases, manifesting a fast electron heating process. As evident in Fig. 1(c), the critical value of $v_D/v_{Te,\parallel}$ ratio that triggers the instability burst is in the range 1.22-1.48; variations are due to the change in $T_{e,\parallel}/T_{i,\parallel}$ and the EVDF.

Fig. 1(d) shows zoom-in on the dynamics of $v_D$ and $v_{Te,\parallel}$ inside a narrow time period $\omega_e t=84000\sim92000$ and demonstrates that there are the slow and fast heating stages of the process (we abbreviate them as SHS and FHS, respectively) during the burst phase. The first stage - SHS lasts for ~1000 electron plasma periods. Later on, the FHS starts when the maximum heating rate becomes much higher. To shed light on the heating mechanisms, we apply the 2D fast Fourier transform (FFT) algorithm to the electric field between $\omega_e t = 86330 \sim 90327$. The obtained wave energy spectrums $|E_\parallel(k, \omega)|^2$, as well as the corresponding fluctuations, are displayed in Figs. 2(a)-2(c) and Figs. 2(d)-2(f), respectively. At the pre-burst phase, two branches of low-frequency (LF) waves in the parallel and anti-parallel directions are observed [Fig. 2(a)]. Both the real frequency $\omega_r$ and wave number $k$ of the LF waves are close to the dispersion relation of the IAWs, indicating that they are the IAWs that occur when the threshold condition of the Bumeman instability is crossed. Besides, a new branch of fast modes which obeys an acoustic-like dispersion relation is also seen [Fig. 2(a)]. These modes propagate in the direction parallel to the electron drift with a speed of $1.08 \times 10^6$ m/s in the ion frame. This value is much higher than the ion acoustic speed $c_s = 2.8 \times 10^4$ m/s, indicating that they are the fast electron modes. We interpret these modes as the nonlinear *electron acoustic waves* (EAWs) for three reasons (similar to findings of Refs. [27-29]): (i) EVDF has the trapped electron population near the phase velocity of EAWs, which makes EAWs not suffer the Landau damping [see Fig. 4(a) therein]; (ii) They have the intermediate frequency between the ion, $\omega_i$, and electron plasma frequencies, $\omega_e$, as is typical for the EAWs; (iii) The value of the wave phase velocity satisfies the plasma dispersion relation calcualted for an EVDF with the plateau [see Supplementary Material (Session B) for a detailed analysis].

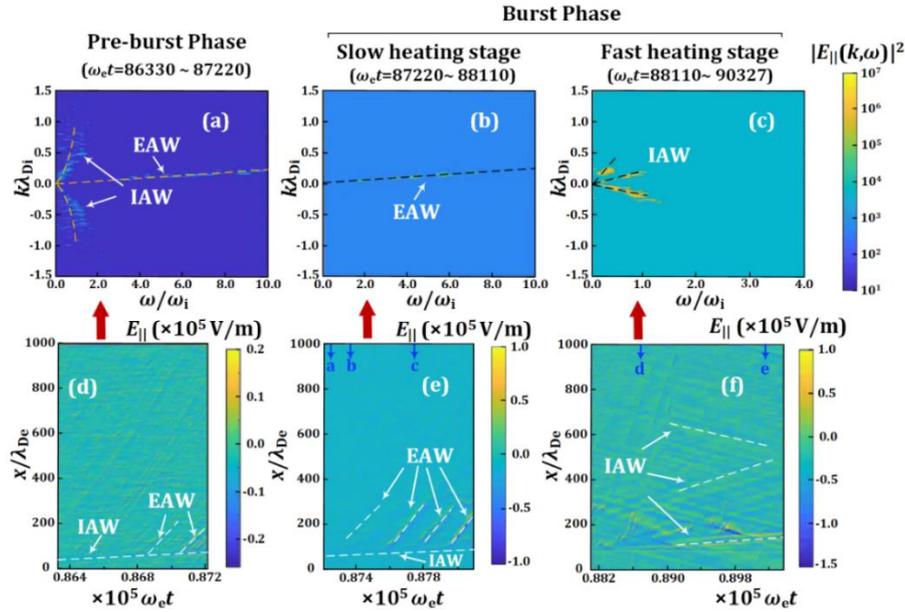

Fig. 2: Evolution of the wave energy density and the electric field fluctuations. (a)~(c) Wave energy spectrum $|E_\parallel(k,\omega)|^2$ and (d)~(f) fluctuations of the $E_\parallel(x,t)$ at the pre-burst phase ($\omega_e t = 86330 \sim 87220$) and the burst phase including the slow heating stage ($\omega_e t = 87220 \sim 88110$) and the fast heating stage ($\omega_e t = 88110 \sim 90327$). $\lambda_{Di}$ is the ion Debye length. The blue arrows

indicate the time moments of the snapshots in Figs. 3(a)-3(e).

In the pre-burst phase, the fluctuations present a turbulent state due to the mixing of three wave branches, but the amplitudes are still too low to trap the electrons [Fig. 2(d)]. After $\omega_e t=87220$, the instability burst occurs [Fig. 2(b)]. During the SHS, the energy density of the EAWs becomes three orders-of-magnitude higher. It can be observed in Fig. 2(e) that the coherent large-amplitude EAWs are emanating from an IAW and propagate along the electron drift direction with a constant speed [Fig. 2(e)]. This indicates that the growth of the EAWs is a result of the nonlinear development of IAWs that modifies EVDF and creates a bump-on tail EVDF [see Fig. 4(a)]. This is somewhat unexpected finding that couples slow ion dynamics to fast electron dynamics. The exact details of the process will require special follow-up studies. During FHS, one can see in Fig. 2(c) that the EAWs decay rapidly while the IAWs grow. The coherent EAWs disappear and the ion-acoustic turbulence emerges instead [Fig. 2(f)].

To show the nonlinear interactions between waves and particles, we plot the snapshots of the electron phase space [Figs. 3(a1)-3(e1)], the ion phase space [Figs. 3(a2)-3(e2)] and the potential profiles [Figs. 3(a3)-3(e3)]. As seen in Fig. 3(a1), at $\omega_e t=87256$, EAWs are still weak while IAW traps electrons and an electron hole forms at $x=90\lambda_{De}$. *Here, we follow the definition in Hutchinson's paper and refer the term electron hole to a localized region where the electron density is lower because of the reduced phase-space density on trapped electron orbits [30].* For better clarification, if the electrons are trapped in an IAW (or EAW) potential and create an EH as the result, we call it "IAW-EH" (or EAW-EH). From Fig. 3(a3) one can see the amplitude of IAW at $x=90\lambda_{De}$ ($e\phi_{max} \approx 10$ eV) becomes higher than $T_e=8.2$ eV. Therefore, the velocity range of trapped electrons has locally extended to $\sim v_{Te} \sim v_{EAW,ph}$ [Fig. 3(a1)], indicating that the resonant regions of IAW and EAWs overlap so that this large-amplitude IAW starts interacting with EAWs. This IAW-EH propagates with the speed $\sim c_s$, indicating that it is a slow electron hole that can reflect and deplete the local ions [Fig. 3(a2)]. Note that the location of this IAW-EH "randomly" changes at different bursts, which may be due to the chaotic evolution of current-driven turbulence [31].

Due to the modification of IAW-EH, the EVDF exhibits a bump-on-tail structure, which triggers the secondary instability and the growth of EAWs. EAW-EHs and the corresponding potential structure can be seen in Figs. 3(b1) and 3(b3). In the meantime, it is observed that IAW-EH starts the coalescence with the adjacent EAW-EH [see the zoom-in view in Fig. 3(f)]. This occurs because the phase areas of the trapped electrons in IAW-EH and EAW-EH have overlapped. Through the coalescence with EAW-EH, IAW-EH grows and further modifies the EVDF, which in turn creates bigger EAW-EHs [Fig. 3(c1)]. The growth of IAW-EH is also manifested by the stronger ion depletion in Figs. 3(b2) and 3(c2). At $\omega_e t=88697$, a giant IAW-EH that traps the main body of electron populations forms. It creates a cavity in the ion phase space and strongly perturbs the nearby ions, thereby generating the IAWs. At $\omega_e t=90158$, the giant IAW-EH has broken and EAW-EHs are not created anymore, while an ion hole is created. At this moment, ion-acoustic turbulence has emerged [Fig. 2(f)] and one can see both electrons and ion density profiles being perturbed in Figs. 3(e1) and 3(e2). For more details, a supplenmetary video showing the evolution of potential, number density, electron and ion phase space can be found in Supplementary Material (Session C) and Supplementary Video.

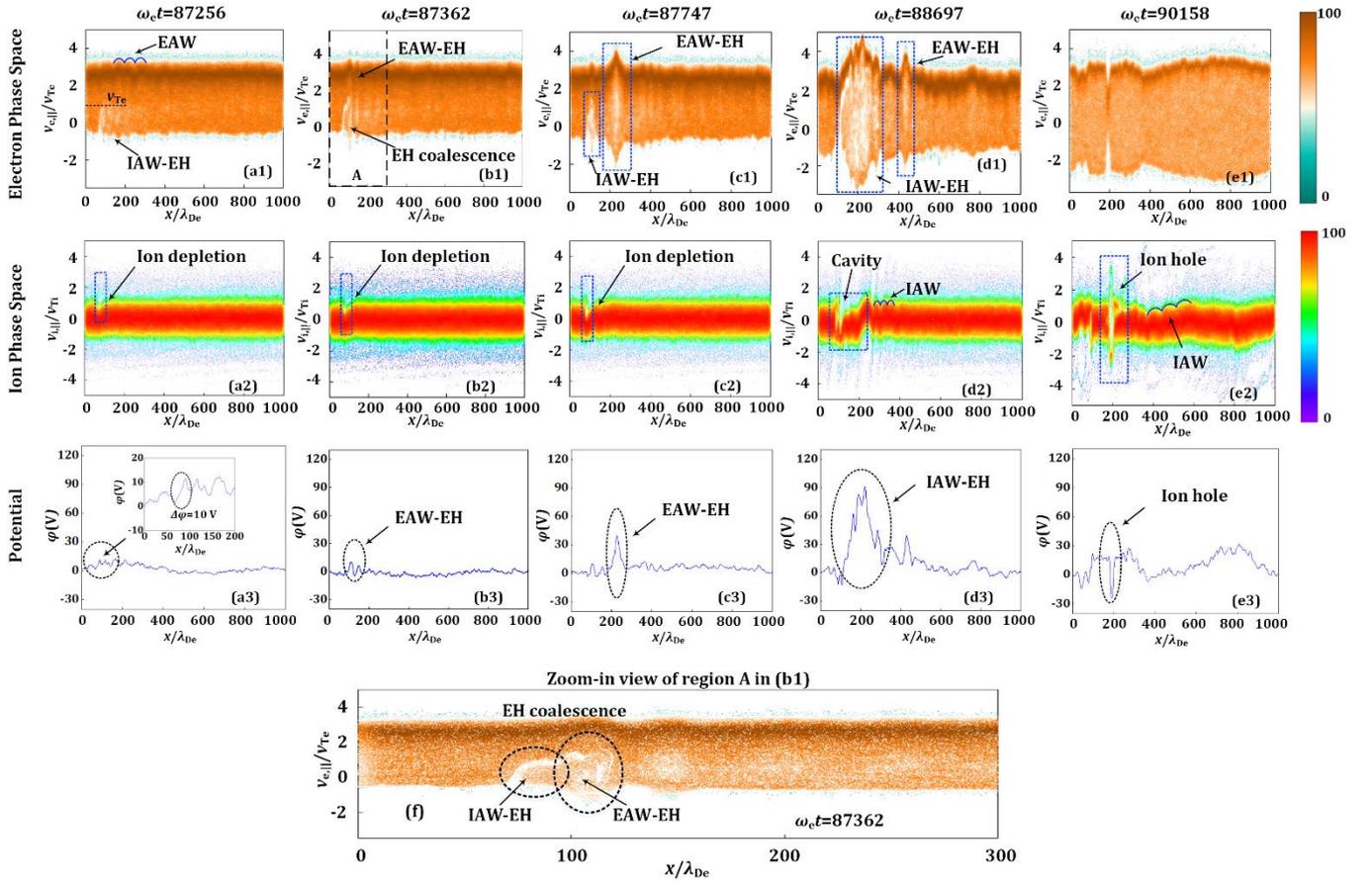

Fig. 3: The snapshots of the electron phase space, the ion phase space, and the potential. The letters "a" to "e" correspond to different moments in time evolution, and the numbers "1" to "3" refer to the electron phase space, the ion phase space, and the potential, respectively. (f) shows the zoom-in view of region A (marked by dash rectangle lines) in (b1). "EH" denotes the "electron hole" formed due to the electron trapping by the large-amplitude waves. The blue wavy lines denote the places where EAW and IAW appear.

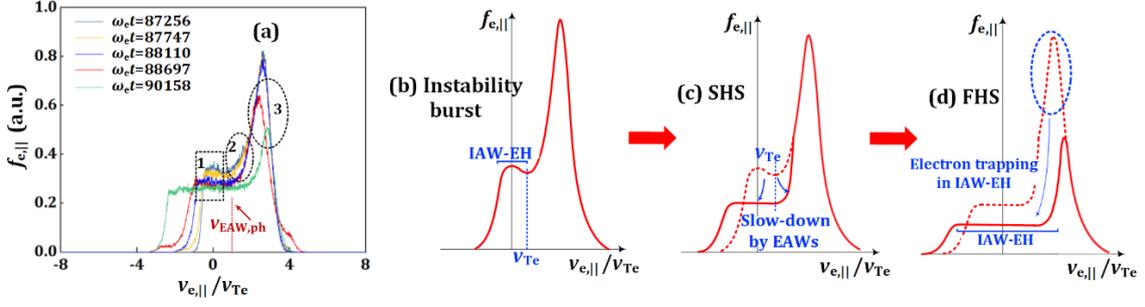

Fig. 4: (a) profiles of EVDF (for all electrons) at $\omega_e t$=87256, 87747, 88110, 88697, and 90158. The first three moments $\omega_e t$=87256, 87747, and 88110 are in the SHS, and the last two $\omega_e t$=88697 and 90158 in the FHS; (b)~(d) Schematics of the variations of EVDF manifesting the electron heating at SHS and FHS. Zones 1,2,3 denote parts of EVDF responsible for different particle-wave interaction processes.

We further analyze how the interplay of the IAWs and EAWs produces electron heating by plotting the EVDF (for all electrons) at different time moments. Correspondingly, the schematics of the variations of EVDF are plotted in Figs. 4(b)-4(d). At the SHS, as one can see from the profiles of EVDF at $\omega_e t$=87256 and 87747 in Fig. 4(a), IAW-EH creates a small bump on the left tail of EVDF (zone 1), thereby generating the EAW-EHs. These EAW-EHs with a relatively high phase velocity drag the electrons with the velocity up to ~$2v_{Te}$ down to the low energy range (zone 2). This is consistent with the fact that the maximum amplitude of EAW is $\varphi_{max}$~$4T_e/e$~33 V [Fig. 3(c3)]. With more and more electrons trapped, fewer free fast electrons can support further growth of EAWs. Instead, the IAW-EH wave becomes giant. At $\omega_e t$=88697, the amplitude of IAW-EH increases up to 90 V [Fig. 3(d3)], signifying that the main body of the drifted electron population (around $3v_{Te}$) is already trapped (zone 3). In this IAW-EH, a strong phase mixing allows the rapid energy exchange between the high- and low-energy electron population, producing a fast electron heating and a huge plateau at $\omega_e t$=90158 [see EVDF Fig. 4(a)].

In summary, we study the nonlinear interplay of electron- and ion-acoustic waves and the resulting electron heating in current-driven turbulence. We show that once $v_D/v_{Te,\|}$ crosses the threshold 1.22-1.48, a slow, large-amplitude ion accoustic wave with a hole in electron phase space (we called it IAW-EH) is generated due to the Buneman instability and creates a bump on the left tail of EVDF. The modified EVDF supports the growth of EAWs and generates fast electron accoustic waves also with holes in the electron phase space. These fast electron waves have large potential amplitude and slow down some high-energy electrons and that leads to electron heating. In the meantime, the IAW-EH grows by the coalescence with other waves by due exchange of trapped electrons in these waves and ultimately becomes large enough to trap the main body of the drifted electron population. In this IAW-EH, electrons are rapidly heated via a strong phase mixing process, thereby creating a plateau in EVDF. Our simulation results demonstrate formation of nonlinear modes with the key features consistent with those measured in current-carrying space plasmas [9, 10], and further manifest that the very large-amplitude ion-acoustic electron holes can be created through the interaction with EAWs and produce very strong electron heating. Similar physical mechanism should also apply to wave generation in plasma discharges with a large current flow (e.g. Ref. [4]).

The work of Jian Chen was partially supported by the China Scholarship Council. The work of Alexander V. Khrabrov and Igor D. Kaganovich was supported by the Princeton Collaborative Research Facility (PCRF) and Laboratory Directed Research & Development (LDRD) projects, which are funded by the U.S. Department of Energy (DOE) under Contract No. DE-AC02-09CH11466. The authors are very grateful to Haomin Sun, Dr. Liang Xu, and Dr. Sarveshwar Sharma for fruitful discussions.